\documentclass[intlimits,twoside,a4paper]{article}
\usepackage[cp1251]{inputenc}

\usepackage[eqsecnum]{cmpj3}

\usepackage{bm}

\issue{2022}{25}{3}{33604}
\doinumber{10.5488/CMP.25.33604}
\title[A polymer-tethered particle confined in a slit]%
{A polymer-tethered particle confined in a slit}
\author[T. Staszewski, M. Bor\'owko]{T. Staszewski\orcid{0000-0002-0284-4253}, M. Bor\'owko\orcid{0000-0003-1461-249X}}
\address{Department of Theoretical Chemistry, Institute of Chemical Sciences, Faculty of Chemistry, Maria Curie-Sk{\l}odowska University in Lublin, Poland}


\date{Received June 24, 2022, in final form September 1, 2022}

\begin{document}

\maketitle

\begin{abstract}
Shape transformations of hairy nanoparticles under confinement are studied using molecular dynamic simulations.
We discuss the behavior of these particles in slits with inert or attractive walls. We assume that only chain-wall interactions are attractive. The impact of the strength of interactions with the walls and the width of the slits on the particle configuration is shown.
In the case of attractive surfaces, we found new structures in which the chains are connected with
both walls and form bridges between them: pillars, symmetrical and asymmetrical spools, and hourglasses. In wide pores with strongly attractive walls, hairy particles adsorb on one of the surfaces and form ``mounds'' or starfish-like strucures.
%
\printkeywords
%
\end{abstract}

\section{Introduction}

Nanoparticles modified with polymeric ligands are a subject of extensive studies since they have a variety of technological applications \cite{1, 2, 3, 4}. Such ``hairy'' particles combine physical and chemical properties of the (inorganic or organic) cores with the features of soft polymeric coatings.
Most of the research concerning the polymer-tethered particles focused on modelling the morphology of polymer canopies by changing ligand properties, the grafting density, the interactions of chains with the environment, and the temperature. The results are summarized in the reviews \cite{1, 2}. The theoretical works were based on the similarity between the polymer-tethered particles and star polymers. Ohno et al. \cite{5} extended the mean-field theory of star polymers developed by Daoud and Cotton \cite{6} to the hairy particles. The methods for the description of the behavior of hairy particles included the self-consistent field model, the scaling theory \cite{7, 8}, the density functional theory \cite{8, 9}, and molecular simulations \cite {10, 11, 12, 13, 14, 15, 16, 17, 18, 19, 20, 21}. Various specific issues regarding the hairy particles were considered, such as the interactions with the surrounding molecules and ions \cite{11, 12}, the reorganization of ligands tethered to nanoparticles under different environmental conditions \cite{13} which results in the formation of ``patchy'' nanoparticles, the change of the particle shape \cite{14, 15} caused by adsorption of small particles, the self-assembly of hairy particles in bulk systems\cite{8, 16, 17, 19, 20, 21}, and many other topics.

Less attention has been paid to the behavior of hairy particles near solid surfaces \cite{22, 23, 24, 25, 26, 27, 28}. The research concentrated mainly on the structure of the thin films of hairy particles deposited on substrates. The film morphology depends on the nature of ligands, surface chemistry, and solvent quality. The presence of a solid surface affects the internal structure of hairy particles. Che et al. \cite{22, 23} studied polystyrene-grafted gold nanoparticles on different substrates and showed that, with increasing polymer-surface interaction energy, the polymer coating of individual particles can spread out to increase its interaction with the surface. Moreover, they studied the influence of the particle density on the structure of the adsorbed film and found strings of particles in the sub-monolayer regime, while in the dense monolayer a well-ordered hexagonal structure was observed. These experimental observations were confirmed by computer simulations \cite{24, 25, 26, 27}. The ``canopies'' of isolated hairy particles near attractive walls are quite similar to the structures formed by star polymers at surfaces \cite{28}.  The mechanism of adsorption of mono-tethered nanoparticles on solid surfaces was also studied using molecular dynamics \cite{29}. Depending on the system parameters, the mono-tethered particles were adsorbed as single particles or as different aggregates and the structure of the adsorbed layer depended mainly on the type of the surface \cite{29}.
Recently, the behavior of ligand-tethered particles confined between two walls was investigated \cite{30, 31}. Ilnytskyi et al.~\cite{30} analyzed the gelation of nanoparticles decorated by liquid-crystalline ligands at the confinement. However, Ventura Rosales et al. \cite{31} showed that the number of patches on particles modified with diblock copolymers can be controlled by tuning the degree of confinement imposed on the particle.

A great number of various commercial products containing nanoparticles create a new type of nanowaste. The adsorption on solids is an effective and safe method for the removal of nanoparticles from the environment \cite{27}. However, there is still a lack of systematic investigations concerning the correlation between the properties of hairy particles and their adsorption on substrates and the morphology of the surface layers.

In this work, we study an idealized coarse-grained model for polymer-tethered particles between parallel walls using molecular dynamics simulations. In general, the behavior of the systems depends on the strengths of interactions between all single entities: cores and segments, as well as their interactions with the substrate \cite{19}. However, simpler models were usually used in the simulations. Usually, purely repulsive cross-interactions mimic incompatibility between cores and polymers. Among the latter models we can distinguish four classes, namely, those with (i) all repulsive interactions \cite{16}, (ii) all attractive interactions~\cite{20}, (iii) the attractive core-core interaction and the repulsive segment-segment interactions~\cite{32}, and inversely, (iv) the repulsive core-core interactions and the attractive segment-segment interactions~\cite{17, 21, 27}. Similarly, interactions of particular ``atoms'' with solid surfaces can be different. The walls can be the same or different (Janus slits). As a consequence, various models of the hairy particles confined in a slit-like pore can be considered.

Our main goal is to study the impact of the confinement in the slit on the shape of an isolated hairy particle. We focus on the ``pure'' effects of the walls. Therefore, we assume that core-core, core-segment, segment-segment, and core-wall interactions are purely repulsive. However, segment-wall interactions can be repulsive or attractive. We show that our model can capture the basic factors determining the behavior of hairy particles in the slits.

\section{Model and simulation method}

We study a single ligand-tethered particle near the solid surfaces. To reduce the number of required ``atoms'', we introduce a coarse-grained model of particles. A single hairy nanoparticle is modelled as a spherical core with attached $f$ chains. Each chain consists of $M$ tangentially jointed spherical segments of identical diameters $\sigma_s$, while the core diameter equals $\sigma_c$. The chain connectivity is enforced by the harmonic segment-segment potentials 
\begin{equation}
u^{(b)}_{ss}=k_{ss}(r-\sigma_s)^2,
\label{eq:ssbonda}
\end{equation}
where $r$ is the distance between segments. The first segment of each chain is permanently tethered to the core at a randomly chosen point on its surface (at the distance $\sigma_{cs} = 0.5(\sigma_c + \sigma_s)$. 

All the ``atoms'' interact via the shifted-force Lennard-Jones potential \cite{33} 
\begin{equation} 
u^{(ij)}=\left\{
\begin{array}{ll}
4 \varepsilon_{ij} \left[ (\sigma_{ij}/r)^{12}-(\sigma_{ij}/r)^6\ \right] + \Delta u^{(ij)}(r), & \ \ \ r < r_{\text{cut}}^{(ij)}, \\
0, & \ \ \ {\rm otherwise,}
\end{array}
\right.
\end{equation}
where
\begin{equation} 
\Delta u^{(ij)}(r)=-(r-r_{\text{cut}}^{(ij)})\partial u^{(ij)}(r_{\text{cut}}^{(ij)}) / \partial r,
\end{equation} 
where $r^{(ij)}_{\text{cut}}$ is the cutoff distance, $\sigma_{ij}=0.5(\sigma_i+\sigma_j)$, $ (i, j = c, s)$, and $\varepsilon_{ij}$ is the parameter characterizing strengths of interactions between spherical species $i$ and $j$. The indices ``c'' and ``s'' correspond to the cores and the chain segments, respectively.

The energy of interactions of segments and core with both surfaces is the
following sum:
\begin{equation}
v_k (z')=v^{(1)} (z') + v^{(2)}(H-z'),
\end{equation} 
while $z'$ is the distance from the surface labeled as ``1'' and $v^{(i)}$, ($i=1, 2$) describes interactions with the $i$-th wall  which are modelled by the 9-3 Lennard-Jones equation  
\begin{equation} 
v_k(z')=\left\{
\begin{array}{ll}
\frac{2}{15} \varepsilon^{(k)}_s \left[ (\sigma_k/z')^{9}-(\sigma_k/z')^3\ \right] , & \ \ \ z '< z_{\text{cut}}^{(k)}, \\
0, & \ \ \ {\rm otherwise,}
\end{array}
\right.
\end{equation}
where $z^{(k)}_{\text{cut}}$ is the cutoff distance, while $\varepsilon_s^{(k)}$ is the parameter characterizing interactions of the $k$-th component with a wall ($k = c, s$). 
In the above, $z$ denotes the distance from the surface, $z^{(k)}_{\text{cut}}$ is the cutoff distance, and $\varepsilon_k $ is the parameter characterizing the interactions of species k with the surface ($k = c, s$). 
To switch on or to switch off the attractive interactions, we use the cutoff distance parameters, at $v_k(z)=0$. 

We assume that only segment-wall interactions can be attractive, while the remaining interactions are softly repulsive. In the last case,  $r^{(ij)}_{\text{cut}} = \sigma_{ij}$. In the framework of the implicit solvent model, our assumptions correspond to the good solvent conditions \cite{9}. The tethered chains are solvophilic and the walls are solvophobic.

The diameter of segments is the distance unit, $\sigma=\sigma_s$, the segment-segment energy parameter, $\varepsilon=\varepsilon_{ss}$ is the energy unit, the mass of a single segment is the mass unity, $m = m_s$, and the basic unit of time is $\tau =\sigma \sqrt{\varepsilon/m}$. We use here the standard reduced quantities, reduced distances $l^*=l/\sigma$, and reduced energies $E^*=E/\varepsilon$. The usual definition of the reduced temperature is introduced, $T^* = k_{\text{B}}T/\varepsilon$, where $k_{\text{B}}$ is the Boltzmann constant. We  also use the reduced densities: the reduced density of cores, $\rho^*_c \sigma_c^3=N \sigma_c^3/V$, the reduced density of segments, $\rho_s^*= NM f \sigma_s^3/V$, where $N$ is the number of particles (cores) and $V$ is the volume of the system.

Molecular dynamics simulations were performed using the LAMMPS package \cite{34, 35} with 
the Nose-Hoover thermostat to regulate the temperature. The simulation protocol is the same as in our previous work \cite{27}. In the simulations, $M = 30$, $f = 30$, $\varepsilon_{ij}^*=1$ ($i,j=c,~s$). The mass of the core is arbitrarily set to $m_c=4m_s$. The energy constants of the binding potentials, $k_{cs}$ and $k_{ss}$, are $1000\varepsilon/\sigma^2$. The reduced temperature is $T^*= 1$.
Moreover,  $\varepsilon^*_s= 1, 3, 6$. These parameters were assumed in the previous works \cite{14, 15, 19, 27}. 

The start configuration was prepared by strong adsorption of the nanoparticle on one of the walls. Then, the other wall was slowly moved to the required distance. Finally, the attractive interactions were turned off and the system was heated. We introduced the desired parameters to such a pre-started system. Each system was equilibrated using at least $10^8$ time steps until its total energy reached a constant level with slight fluctuations around a mean value. The production runs were for at least $10^7$ time steps.

Examples of the configurations are presented using the OVITO \cite{36}.

\section{Results and discussion}
\subsection{Internal structure of hairy particles in slits}

\begin{figure*}
\centering
\includegraphics[width=4.5cm]{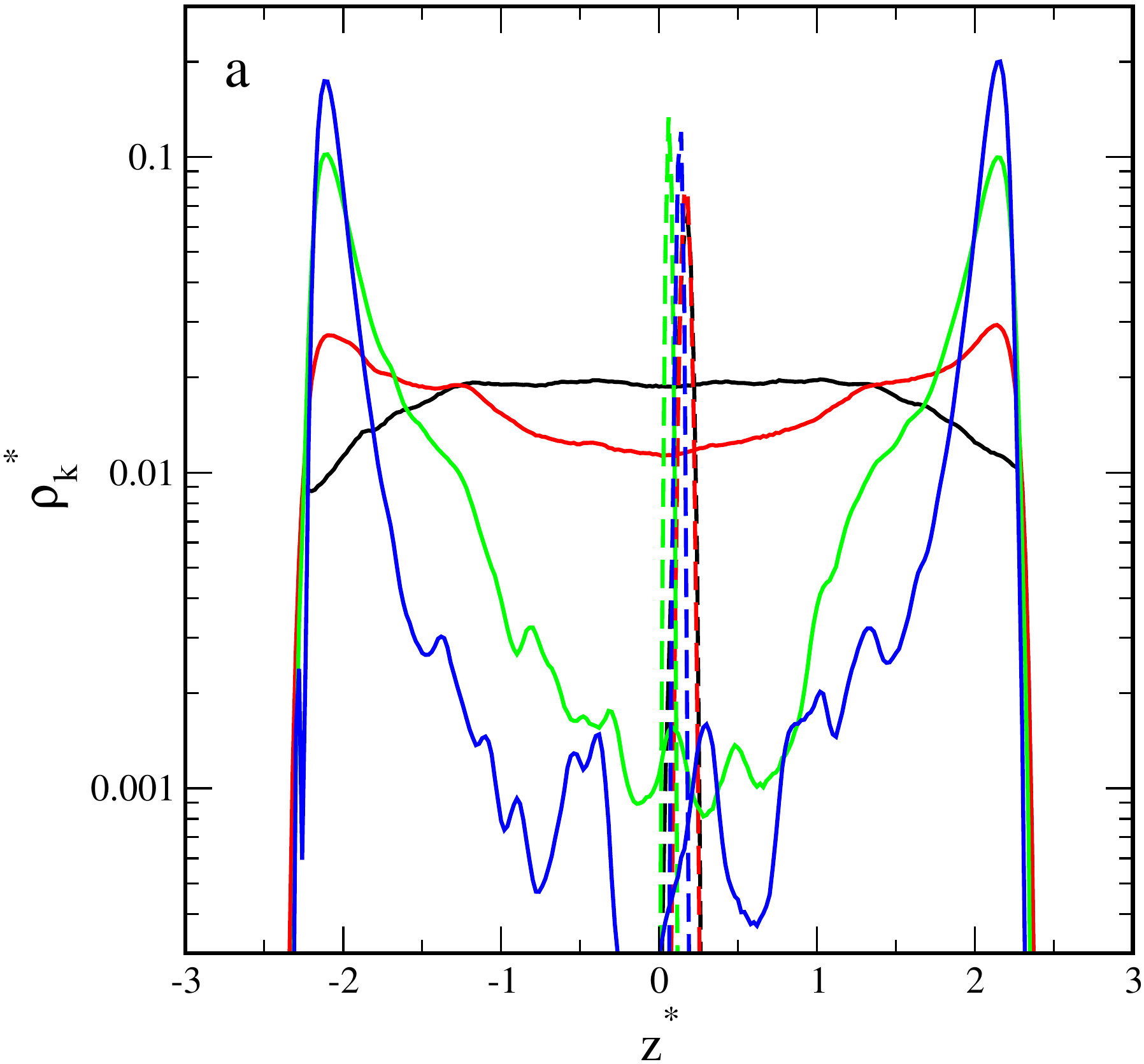} \includegraphics[width=4.5cm]{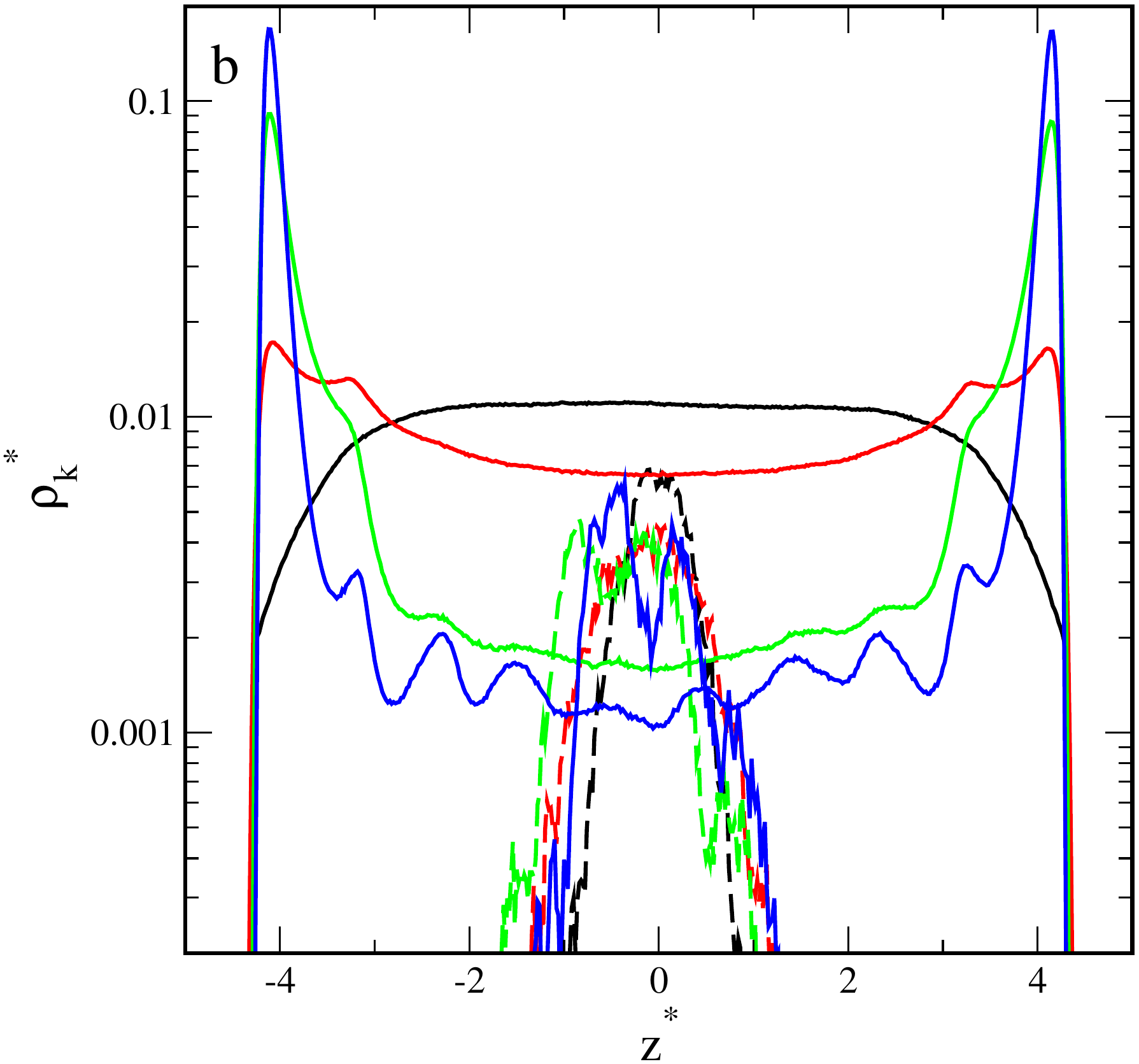} \includegraphics[width=4.5cm]{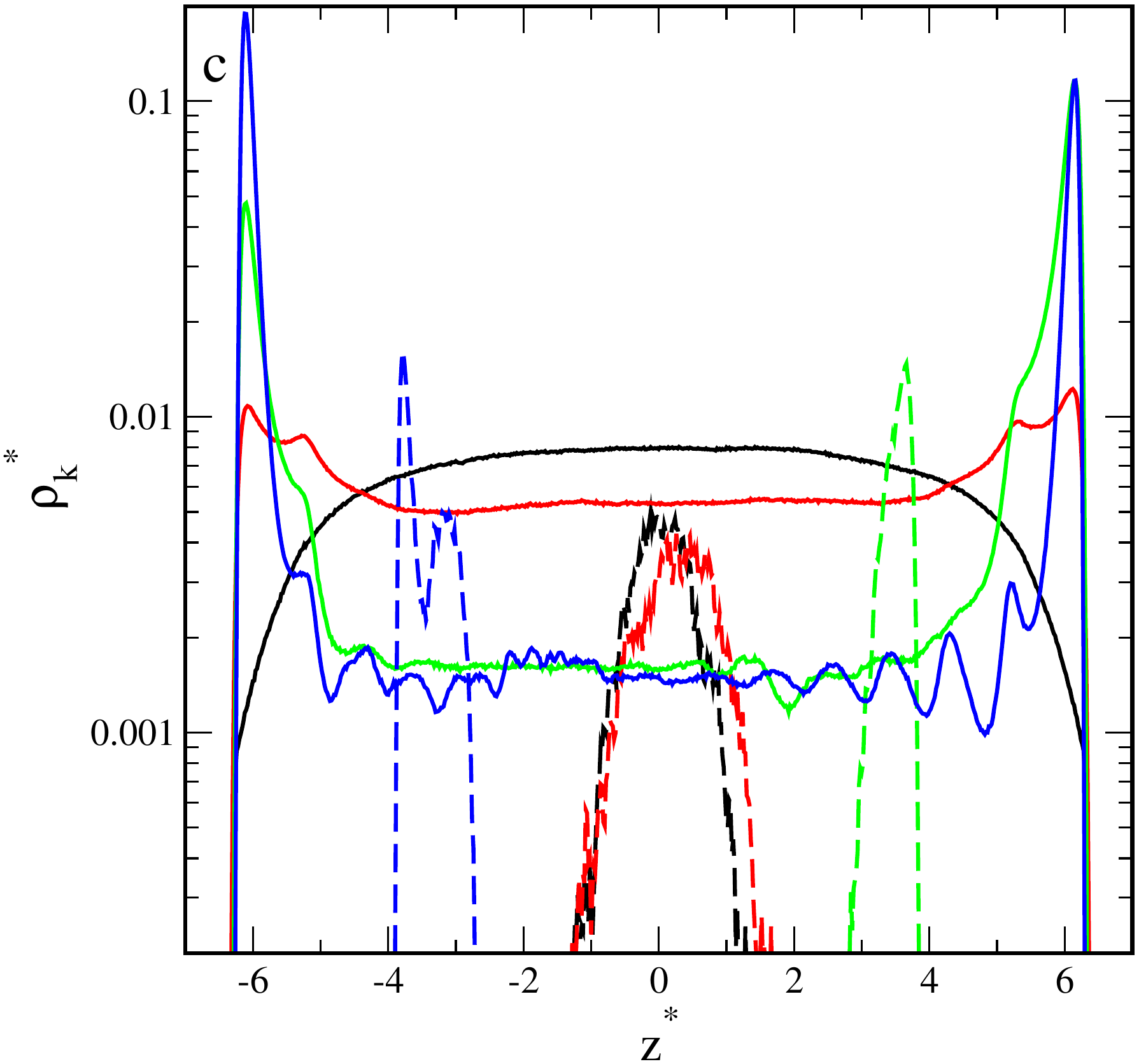} \\
\includegraphics[width=4.5cm]{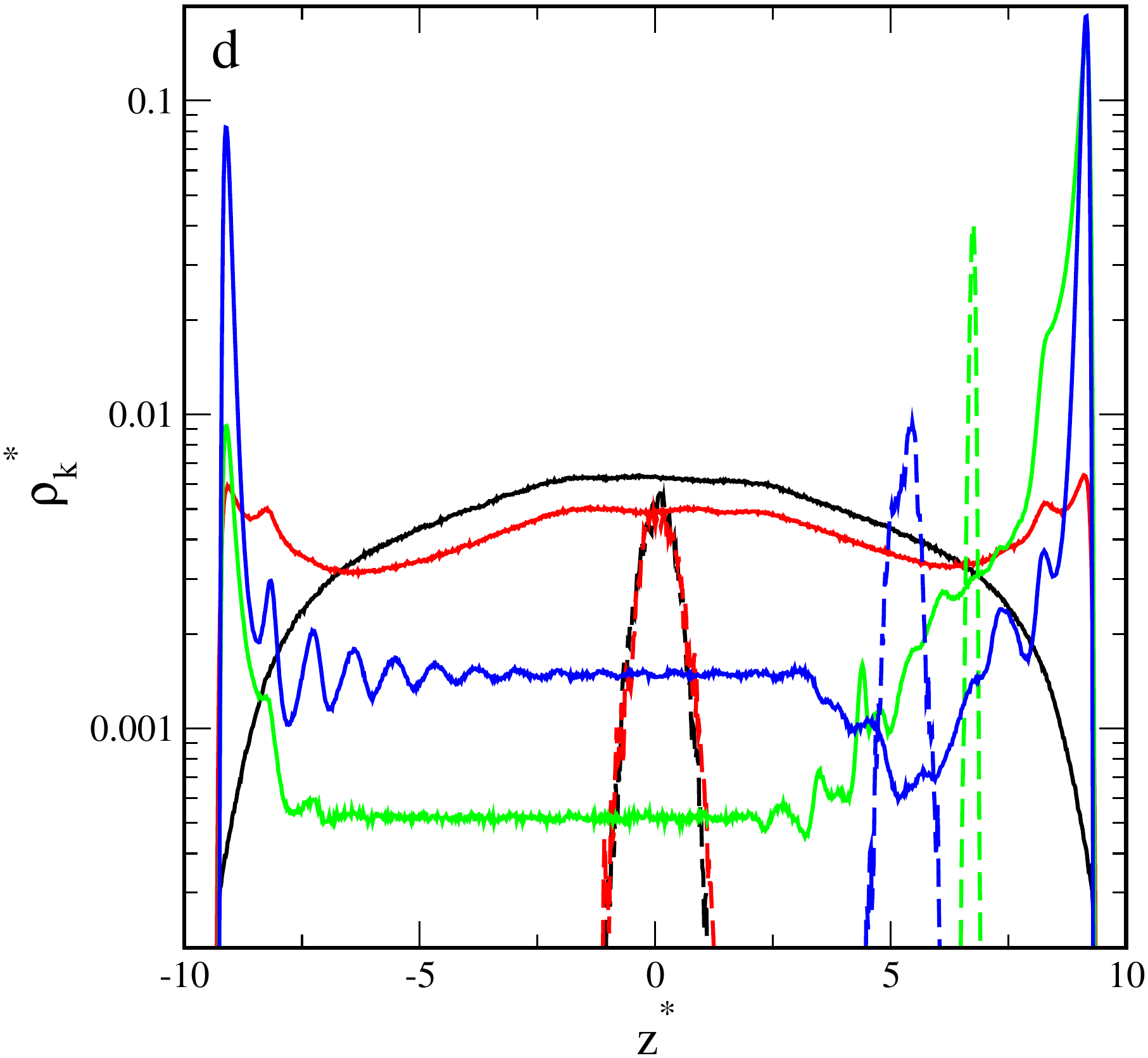} \includegraphics[width=4.5cm]{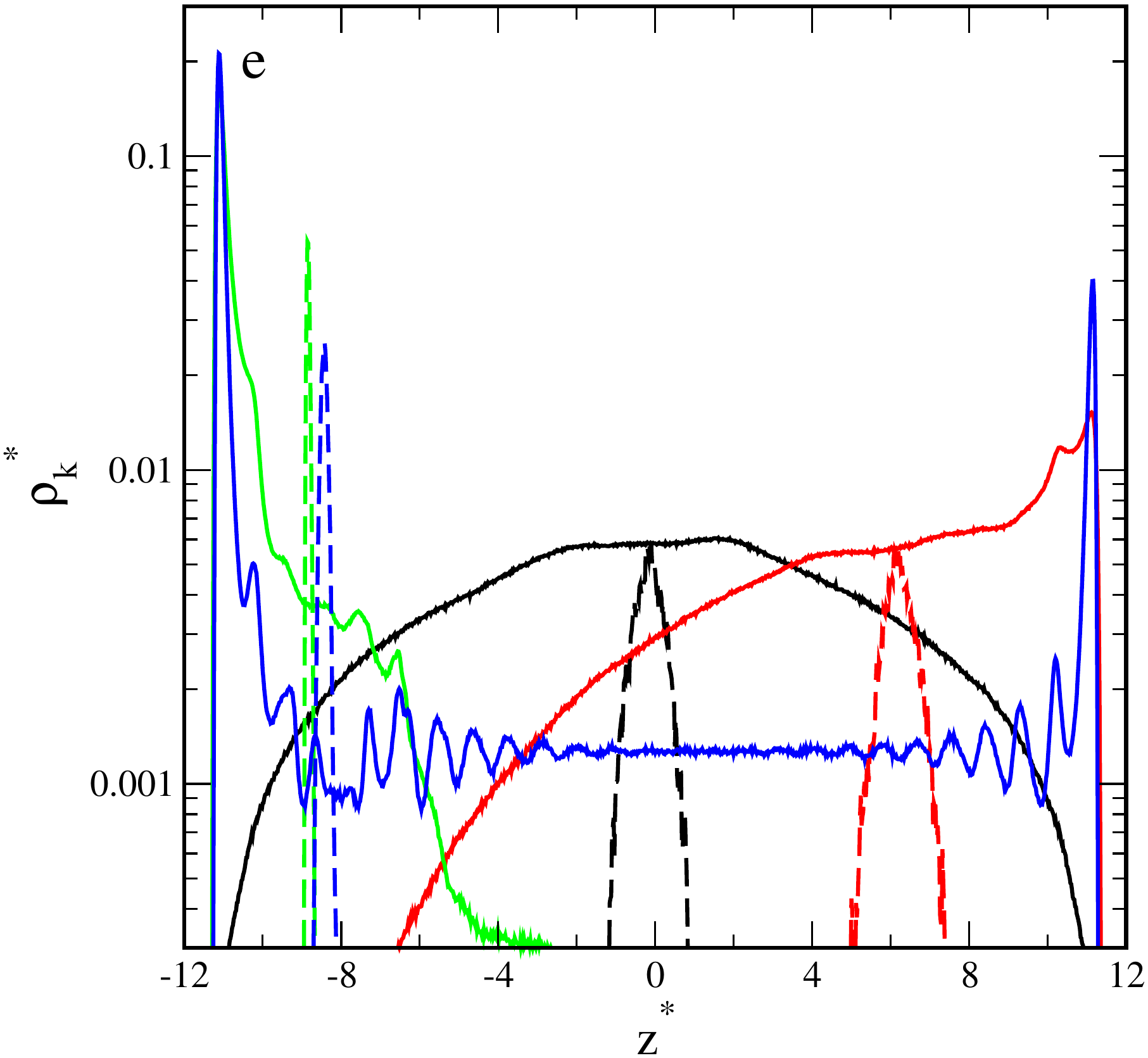} \includegraphics[width=4.5cm]{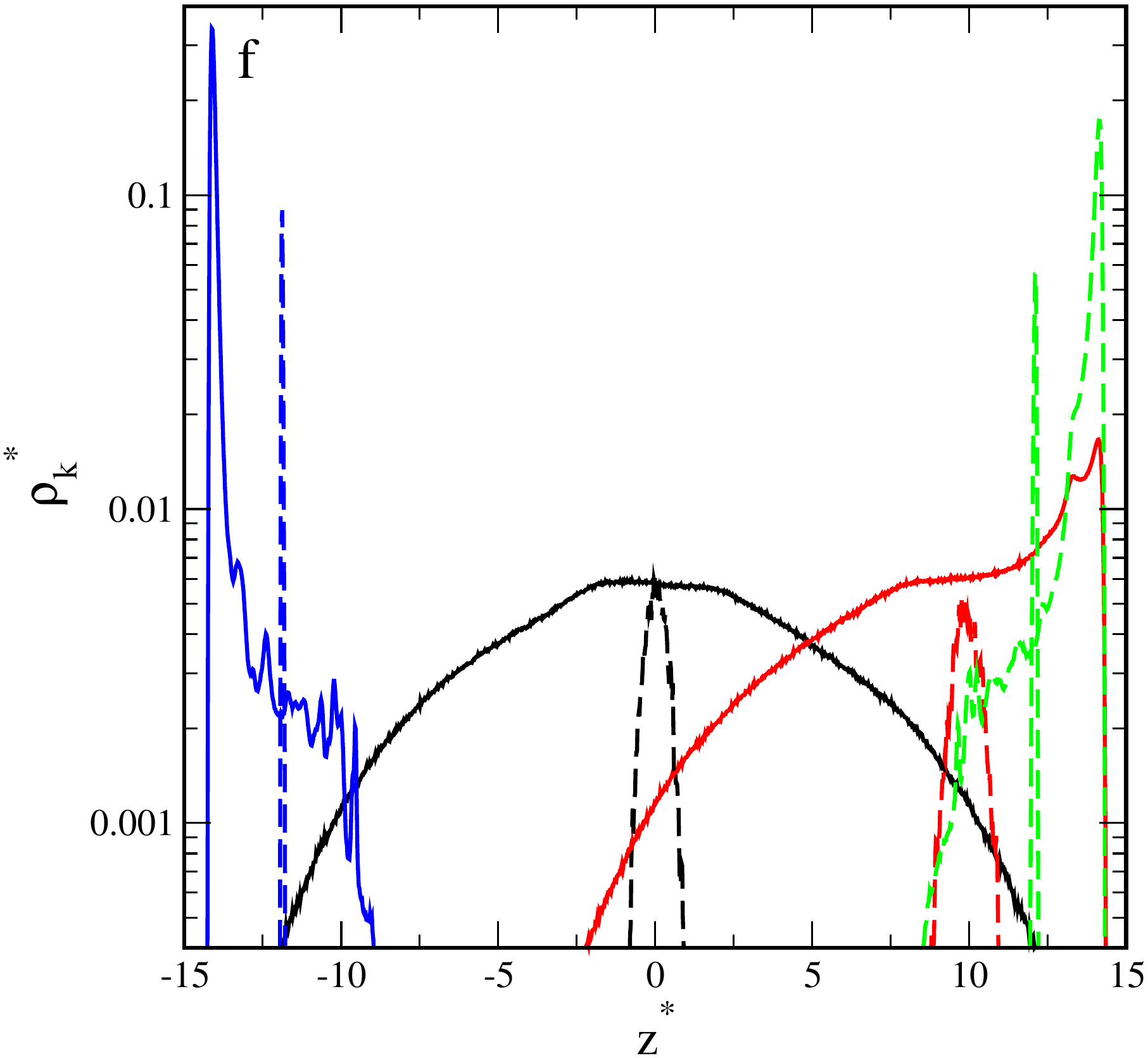}
\caption{(Colour online) Density profiles of the chain segments ($k=s$, solid lines) and the cores ($k=c$, dashed lines)
 in the slits with repulsive walls (black lines) and attractive walls with different parameters $\varepsilon^*_s$: 1 (red lines), 3 (green lines), and 6 (blue lines), and different wall separations: (a) $H^*=6$, (b) $H^*=10$, (c) $H^*=14$, (d) $H^*=20$, (e) $H^*=24$, and (e) $H^*=30$. The ordinates are scalled logarithmically. }
\label{fig1}
\end{figure*}

We begin with the analysis of the structure of a hairy particle confined between two walls. Different wall separations are considered: $H^*=7, 10, 14, 20, 24, 30$. In figure~\ref{fig1} we have plotted the density profiles of segments, $\rho_s^*(z^*)$ (solid lines), and the density profiles of cores, $\rho_c^*(z^*)$ (dashed lines). We show here the density profiles for inert walls (black lines) and attractive walls of different strengths: $\varepsilon_s^*=1$ (red lines), $\varepsilon_s^*=3$ (green lines), and $\varepsilon_s^*=6$ (blue lines). Note that the density is plotted in a logarithmic scale. 

Firstly, we analyse the distributions of cores in the slits. In all cases, the core density profiles have one symmetric peak.  A wider peak reflects stronger oscillations of the core around its average position. In general, oscillations are stronger for repulsive and weakly attractive walls, while the narrow peaks are visible for highly adsorbing surfaces. However, it is interesting that for $H^*=10$ and $H^*=14$, considerable oscillations are observed also for strongly attractive walls. 

\begin{figure}
\centering
\includegraphics[width=6.0cm]{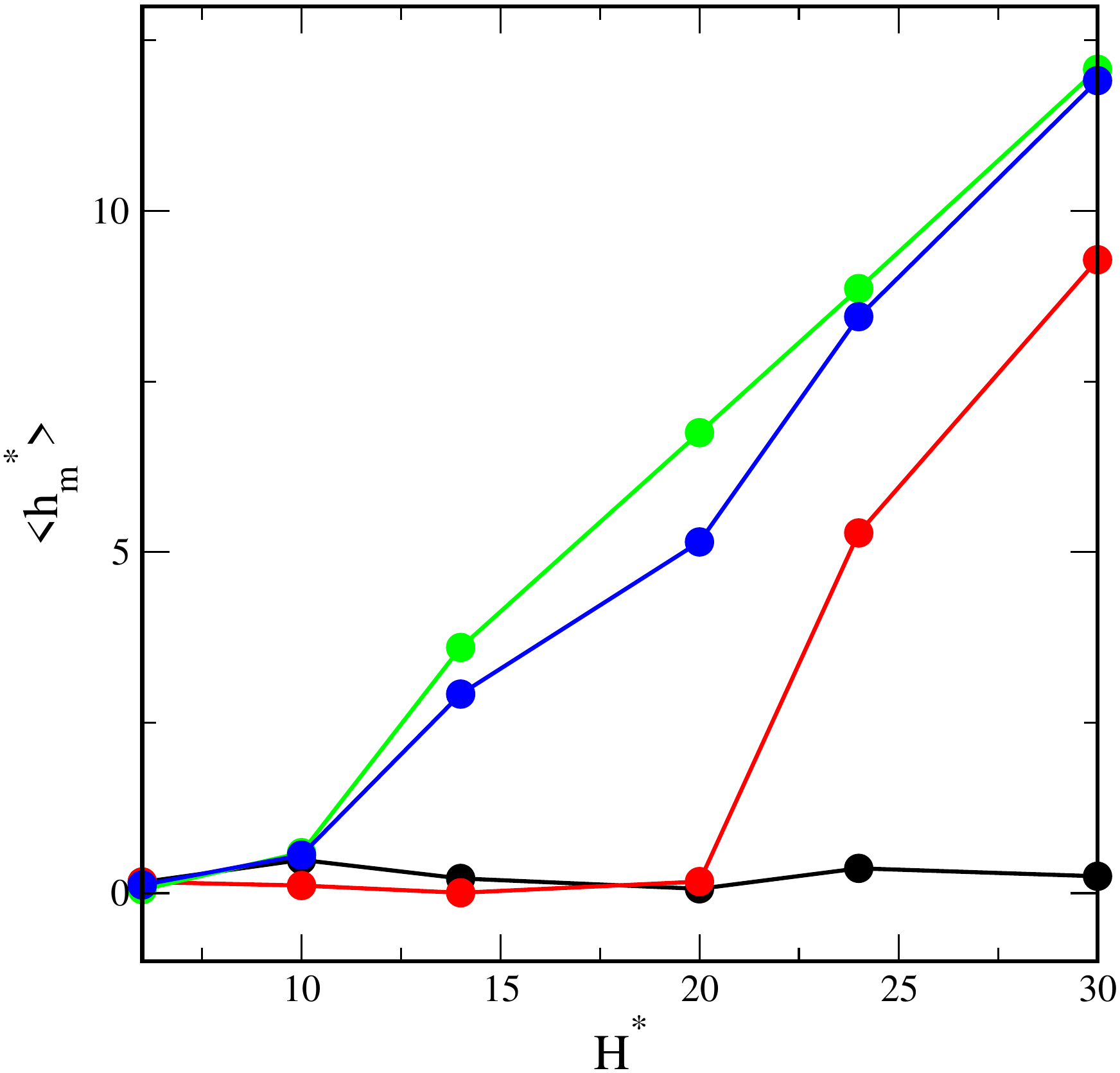}
\caption{(Colour online) The average distance of the core from the center of the slit as the function of the pore width for repulsive walls (black lines) and attractive walls with different energy parameters $\varepsilon^*_s$: 1 (red lines), 3 (green lines), and 6 (blue lines).}
\label{fig2}
\end{figure}

The average distances of the core from the middle of the pore, $h_m^*$, are shown in figure~\ref{fig2}. For inert walls, the core is always located of the slit center,  $h_m^* \approx 0 $. In the case of attractive walls, however, the plots $h_m^*$  vs $H^*$ are completely different. Initially,  $h_m^* \approx 0 $ and after reaching a certain threshold width, the $h_m^*$ increases almost linearly with the pore width. The $h^*_m$ starts to increase at $H^*=20$ for $\varepsilon_s^*=1$ and at $H^*=10$ for strongly attractive walls. In such pores, the core  approaches one of the walls.

Now, we focus on the analysis of the segment density profiles, $\rho_s^*(z^*)$. We checked that all profiles in the $xy$-plane are symmetrical. In the case of repulsive walls, the segment density is lower at the proximity of the walls, and it is higher near the center. With an increase of the  width of the pore, the minima near the walls become deeper, while one maximum appears at the pore center. This indicates that the polymer canopy becomes more spherical. Other segment distributions are found for the weakly attractive walls ($\varepsilon_s^*=1$). In narrow pores (figure~\ref{fig1}a,b,c,d), the segment density increases in the immediate proximity of the walls and is quite high elsewhere. With increasing the pore width, the peaks at both walls become lower. However, the shape of the density profile at the central part of the system depends on the wall separation. For very narrow slits ($H^*=6, 10$), there is a shallow minimum, for $H^*=14$, we see a plateau, while a low wide maximum is found for $H^*=20$. In the wide slits, the particle adsorbs on one wall (figure~\ref{fig1}e,f). In this case, the core is located close to this wall. The segment density has a maximum near the surface and gradually decreases.

\begin{figure*} 
\centering
\includegraphics[width=12cm]{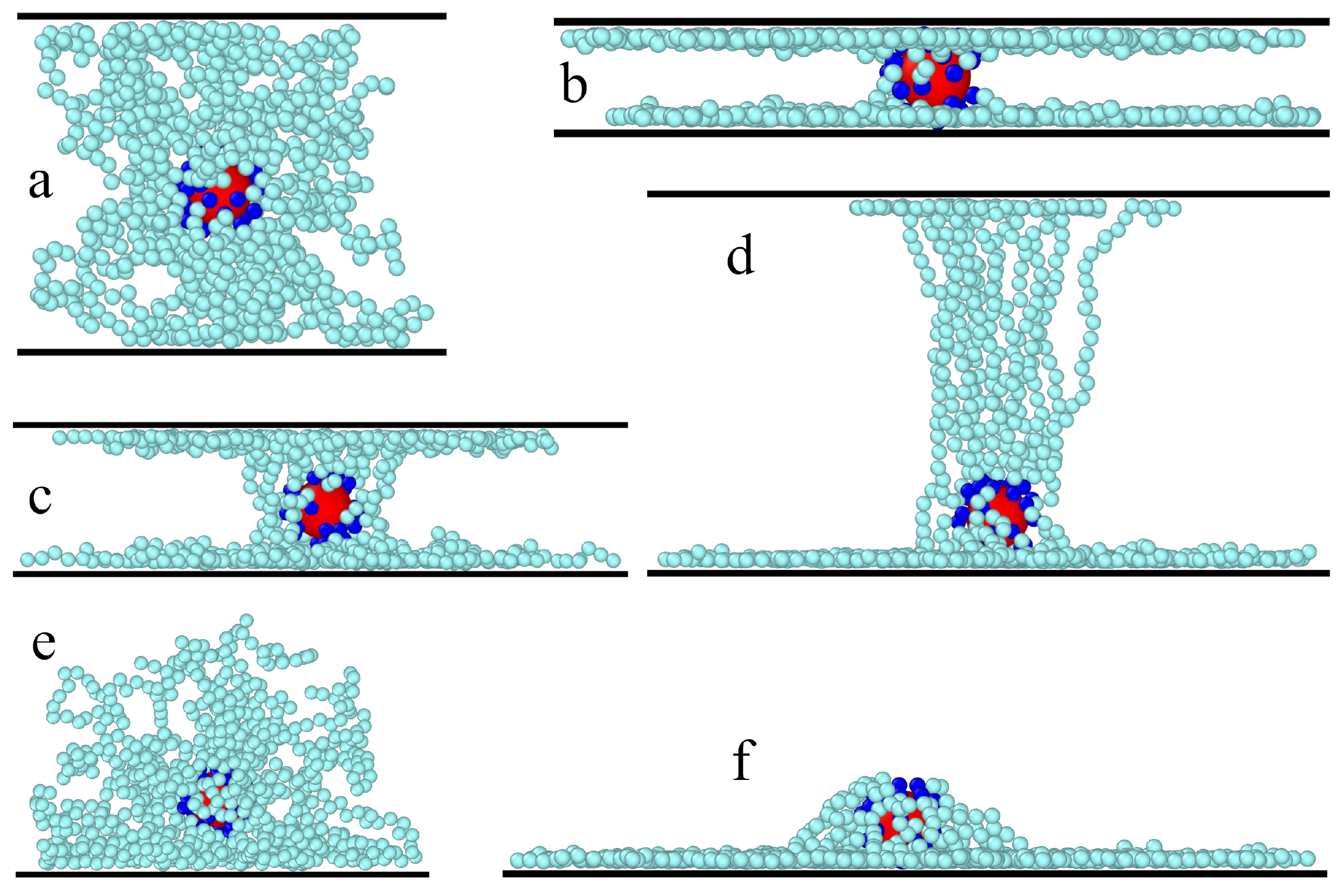}
\caption{(Colour online) Examples of the equilibrium configurations of hairy particles with fixed ligands in different slits. (a) $H^*=20$, $\varepsilon_s^*=1$ --- pillar (P); (b) $H^*=6$, $\varepsilon_s^*=6$ --- hourglass (H); (c) $H^*=10$, $\varepsilon_s^*=3$ --- symmetrical spool (S); (d) $H^*=24$, $\varepsilon_s^*=6$ --- asymmetrical spool (S1); (e) $H^*=30$, $\varepsilon_s^*=1$ --- mound (M); (f) $H^*=30$, $\varepsilon_s^*=6$ --- flattened mound (M1). The red sphere represents the core, and navy blue spheres correspond to bonding segments, light blue spheres represent the remaining segments.}
\label{fig3}
\end{figure*}

Let us discuss the results obtained for slits with very attractive walls. In the case of the narrowest slit, the segment profiles have very high peaks at surfaces and decrease to deep minima at $z^*=0$. For $H^*=10, 14, 20$, one can see a plateau in the central part of the system. In the case of the widest slits (figure~\ref{fig3}e,f), the hairy particle adsorbs on one of the walls. For a more attractive surface, the peak at the wall becomes higher and narrower.

Based on the above-discussed density profiles of segments and the distance of the core from the center of the slit we have classified the configurations of a hairy particle. In the slits with repulsive walls, the segment clouds are flattened spheroids as was shown by Ventura Rosales et al. \cite{31}. Much more interesting results were obtained for attractive walls. The most representative examples of these configurations are shown in figure~\ref{fig3}. We distinguish two basic conformations of hairy particles: bridges and mounds. In the first case, the chains are connected with both walls and, together with the core, form a bridge between them (parts a-d), while the mounds are adsorbed on one of the walls (parts e, f). We  also introduced a more detailed classification. We observe four types of bridges, namely, symmetrical spools~(S), asymmetrical spools~(S1), hourglass (H), and pillars (P). For the spools, the density of segments has high and sharp peaks near the walls and is quite high elsewhere. In the case of symmetric spools, the $h^*_m \approx 0$ and their flanges are almost the same. However, for asymmetrical spools, the core lies closer to one of the walls and the flanges are different. If the density of segments is very low at the center of the pore and gradually increases to maximum values near the walls, we classify such a conformation as an hourglass. In the case of pillars, the density of segments is almost the same in the whole slit, with relatively low peaks at the surfaces. For an hourglass-like structure, however, the density of segments is very low in the  middle of the slit and gradually increases reaching the maxima at walls. Among mound-like conformations, we can distinguish a mound (M) and a flattened mound (M1). The latter one corresponds to a starfish configuration of adsorbed star polymers \cite{28}. 

\subsection{Geometrical properties}

Various parameters can characterize the geometry of hairy particles\cite{37}. One of such characteristics can be the radius of gyration of the cloud of segments
\begin{equation}
R_g^2= \frac{1}{N'} \Big \langle \sum_{i=1}^{N'} {\textbf{r}_{i0}^2 } \Big \rangle,
\label{eq3.1}
\end{equation}
where $\textbf{r}_{i0}=\textbf{r}_i-\textbf{r}_0$, and $\textbf{r}_i$ and $\textbf{r}_0$ are positions of the ith segment and the center of mass, respectively, and $N' = fM$.

 The vectors $\textbf{r}_{i0}$ in equation~(\ref{eq3.1}) can be resolved into components parallel to the axes $x$, $y$, $z$ and can be used to calculate the corresponding radii of gyration labeled $R_{g \alpha}^2$ ($\alpha=x,y,z$). We carried out simulations for bulk systems and found $R_{g0}=9.18$. We  checked that 
 $R_{gx0}^2=R_{gy0 }^2=R_{gz0}^2=R_{g0}^2/3$. We  also calculated the average of components of the radius of gyration in directions parallel to the walls, defined as $R^2_{gxy}= 0.5(R^2_{gx}/R^2_{gx0} + R^2_y/R^2_{gy0})$. Notice that all the radii of gyration are divided by their bulk counterparts.

\begin{figure}[!t]
\centering
\includegraphics[width=7.0cm]{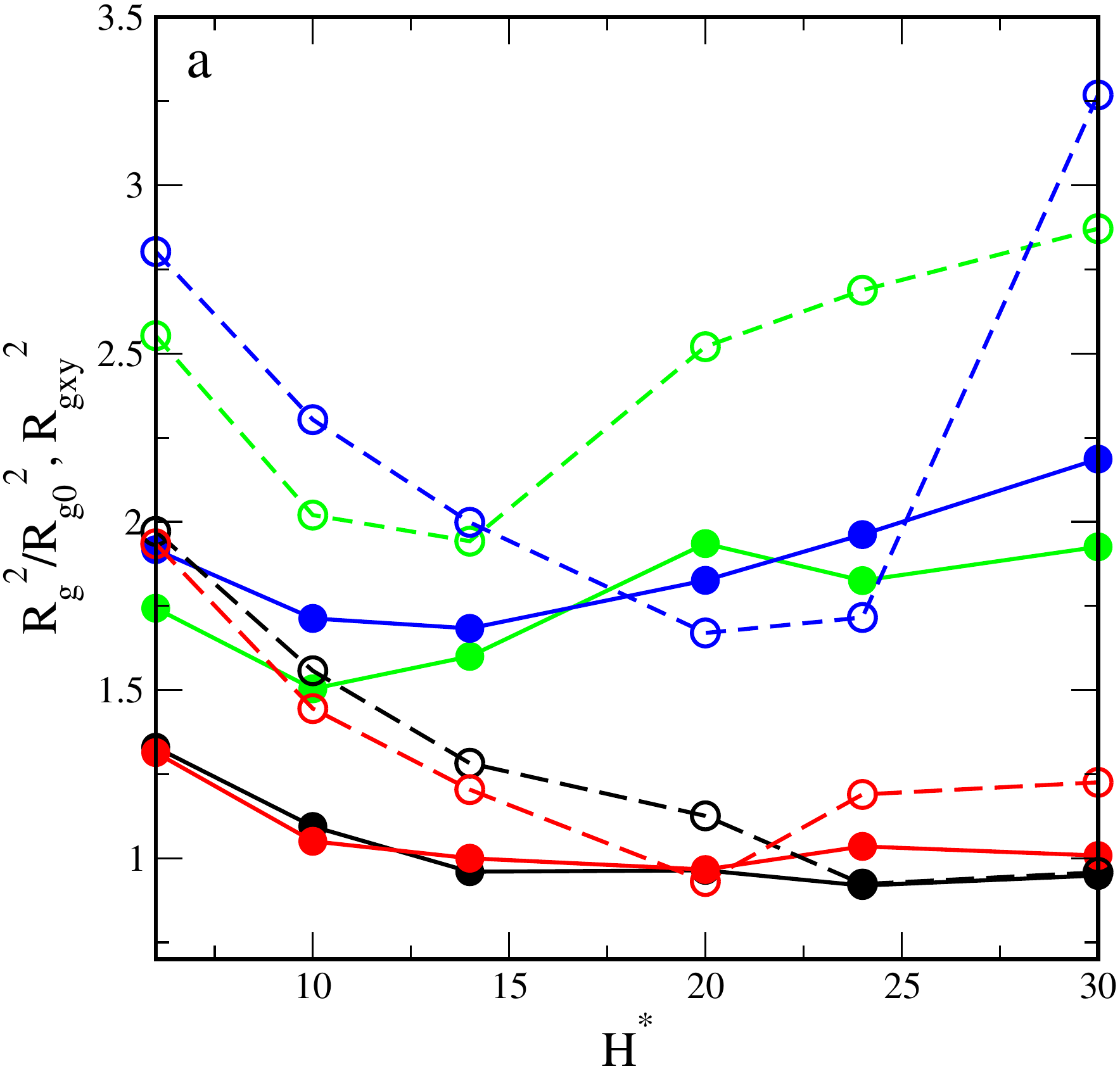} \includegraphics[width=7.0cm]{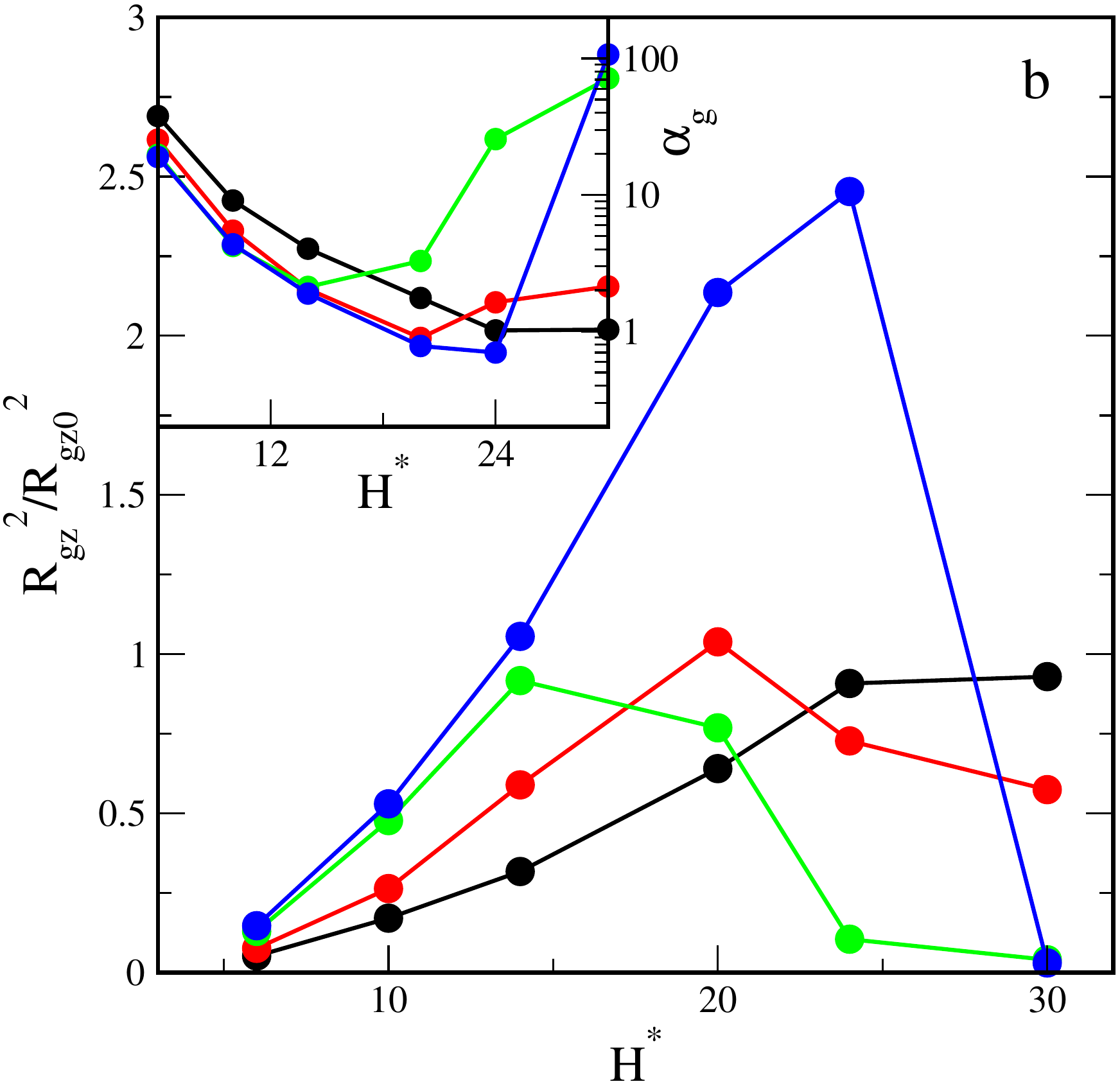}\\
\caption{(Colour online) (a) The squared radius of gyrations (solid lines) and the averages of the components of the squared radius of gyration in directions parallel to the walls (dashed lines); (b) The $z$-th component of the squared radius of gyration and the ratio $\alpha_g$ (inset). All the radii of gyration are divided by their bulk counterparts. These observables are plotted as functions of the wall separation, for repulsive walls (black lines) and attractive walls with different energy parameters $\varepsilon^*_s$: 1 (red lines), 3 (green lines), and 6 (blue lines).}
\label{fig4}
\end{figure}

In figure~\ref{fig4}a we plotted the radius of gyration (solid lines) and the average of components of the radius of gyration in directions parallel to the walls (dashed lines), as functions of the  width ot the pore. 
In the case of repulsive and weakly attractive walls, with increasing $H^*$, the total radius of gyration slowly decreases to the bulk values ($R^2_g/R^2_{g0}$ tends to 1). For strongly attractive walls, the total radii of gyration are several times higher. Moreover, the radius of gyration decreases to a minimum at $H^*=14$. With a further increase in $H^*$, the radius of gyration gradually increases for $\varepsilon^*_s=6$, while for $\varepsilon^*_s=3$, after an initial increase, it remains almost unchanged. In narrow slits, the radius of gyration increases as segment-wall interactions become stronger, although the opposite effect is observed in the case of the widest pores.
The average of components of the radius of gyration in directions parallel to the walls changes with increasing $H^*$ in a similar way. However, the minima are shifted to higher values of $H^*$.
In figure~\ref{fig4}b we show the dependence of the component $R^2_{gz}$ on the  width of the pore.
 For the repulsive walls, the $R^2_{gz}/R^2_{gz0}$ monotonously increases to unity. In the case of attractive walls, these functions have maxima at $H^*=20$ ($\varepsilon^*_s=1$), $H^*=14$ ($\varepsilon^*_s=3$) and $H^*=24$ ($\varepsilon^*_s=6$). The maxima correspond to points of transformations to the mound-like structures ($\varepsilon^*_s=1, 6$) or a more asymmetrical spool ($\varepsilon^*_s=3$).

The shape of the hairy particle can be also described by the ratio $\alpha_g^2=R^2_{gxy}/R^2_z$ (the inset in figure~\ref{fig4}b). We see here that in most systems, the chains are much more extended in the $xy$-plane ($\alpha_g> 1$). Of course, for the repulsive surfaces, $\alpha_g$ smoothly tends to unity as $H^*$ rises. It is interesting, however, that for the ``strongest walls'' and $H^* = 20, 24$, this ratio is also  close to unity. In this case, the diameters of ``collars'' of spools adsorbed on the walls are similar to
the pore widths. 

The shape of the cloud of segments can be characterized by the shape parameters which are defined
using the gyration tensor \cite{37}
\begin{equation}
G_{\alpha \beta}= \frac{1}{N'} \Big \langle \sum_{i=1}^{N'} {(r_{i,\alpha}-r_{0,\alpha})(r_{i,\beta}-r_{0,\beta})} \Big \rangle,
\end{equation}
where $r_{i,\alpha}$ and $r_{0,\alpha}$ are the $\alpha$ component ($\alpha, \beta=x, y, z$ ) of the $i$-th segment and of the center of mass of the segment cloud, respectively. 

By diagonalization of the gyration tensor, one can obtain its eigenvalues $\lambda_i$ ($i = 1, 2, 3$), where $\lambda_1 \geqslant \lambda_2 \geqslant \lambda_3$. Hence, we get three invariants: $I_1=\lambda_1+\lambda_2+\lambda_3$, $I_2=\lambda_1\lambda_2+\lambda_1\lambda_3+\lambda_2\lambda_3$, and $I_3=\lambda_1\lambda_3\lambda_3$. The invariants are used to define the shape parameters, the relative shape anisotropy, prolateness, and the acylindricity \cite{31, 37}. 
The relative shape anisotropy is given by
\begin{equation}
\kappa=1-3 \langle I_2/I^2_1 \rangle.
\end{equation}
The relative shape anisotropy varies from 0 to 1. In the case of perfectly spherical objects, $\kappa=0$. However, for linear rigid rods $\kappa=1$. For a regular planar array $\kappa=0.25$ \cite{37}.

The prolateness is defined by
\begin{equation}
S= \langle (3\lambda_1-I_1)(3\lambda_2-I_1)(3\lambda_3-I_1)/I^3_1\rangle.
\end{equation}

The prolateness takes values in the range from $-0.25$ to 2. The negative values correspond to oblate shapes, while the positive ones characterize prolate shapes. In the case of perfectly oblate clouds ($\lambda_1=\lambda_2>\lambda_3$). In perfectly prolate clouds, the shorter axes of the elipsoid are the same ($\lambda_1>\lambda_2=\lambda_3$)~\cite{38}.

The acylindricity can be expressed as
\begin{equation}
C= \langle (\lambda_2-\lambda_3)/I_1 \rangle,
\end{equation}
where $C \geqslant 0$ and $C = 0$ describes a perfect cylindrical symmetry. In the above equations, $\langle ... \rangle$ denotes an average over all configurations.

\begin{figure}[!t]
\centering
\includegraphics[width=4.5cm]{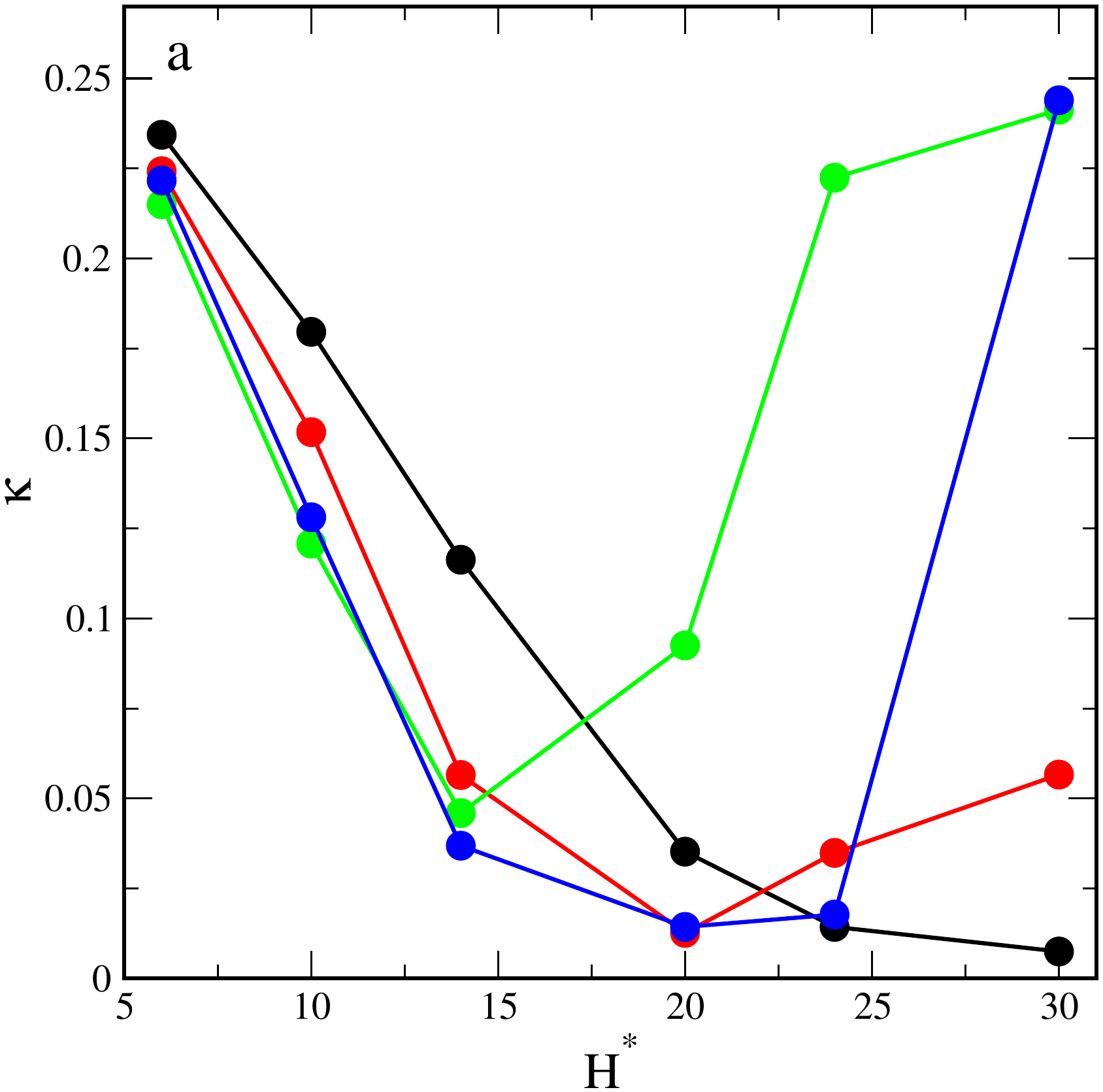} \includegraphics[width=4.5cm]{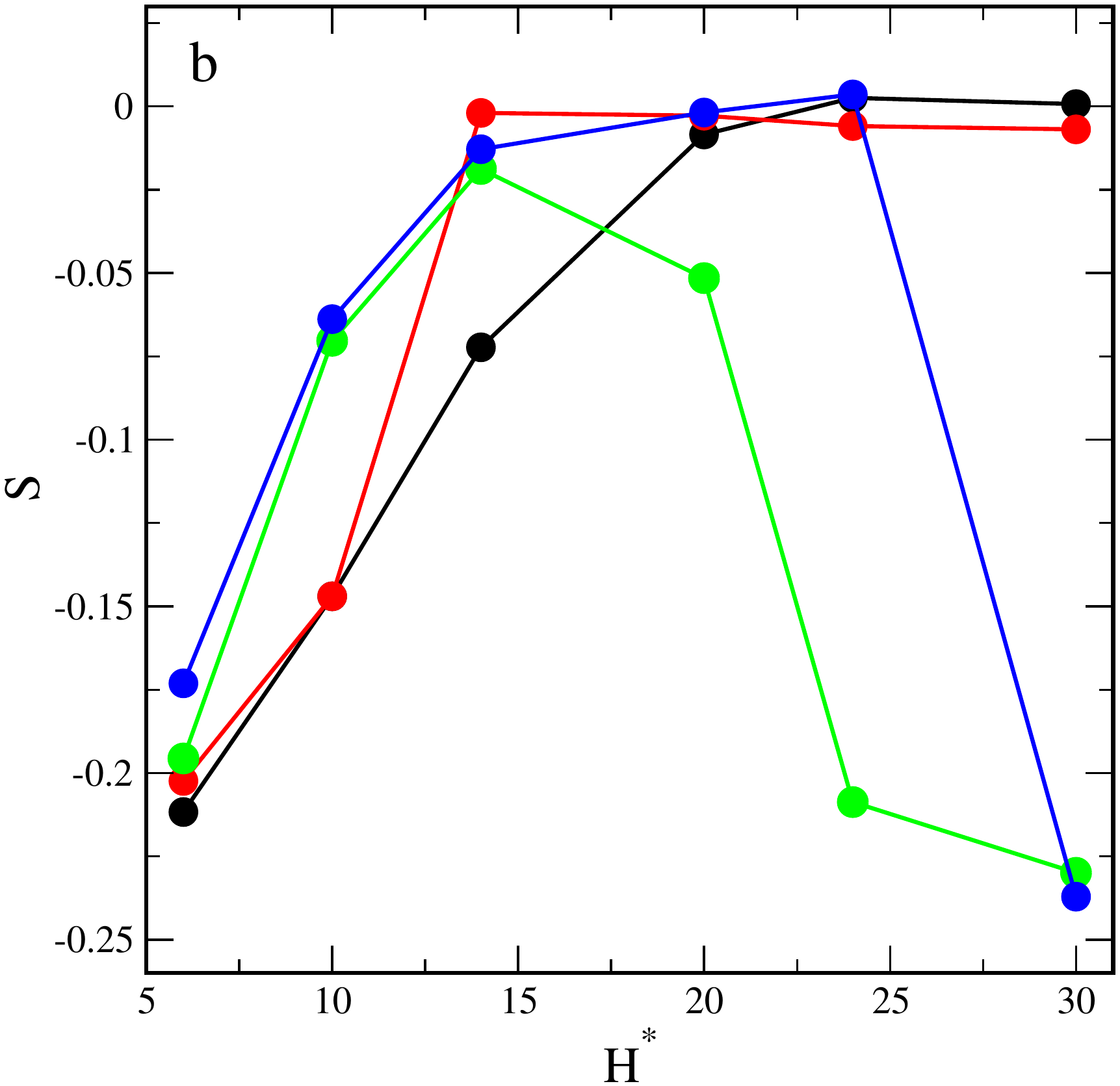} \includegraphics[width=4.5cm]{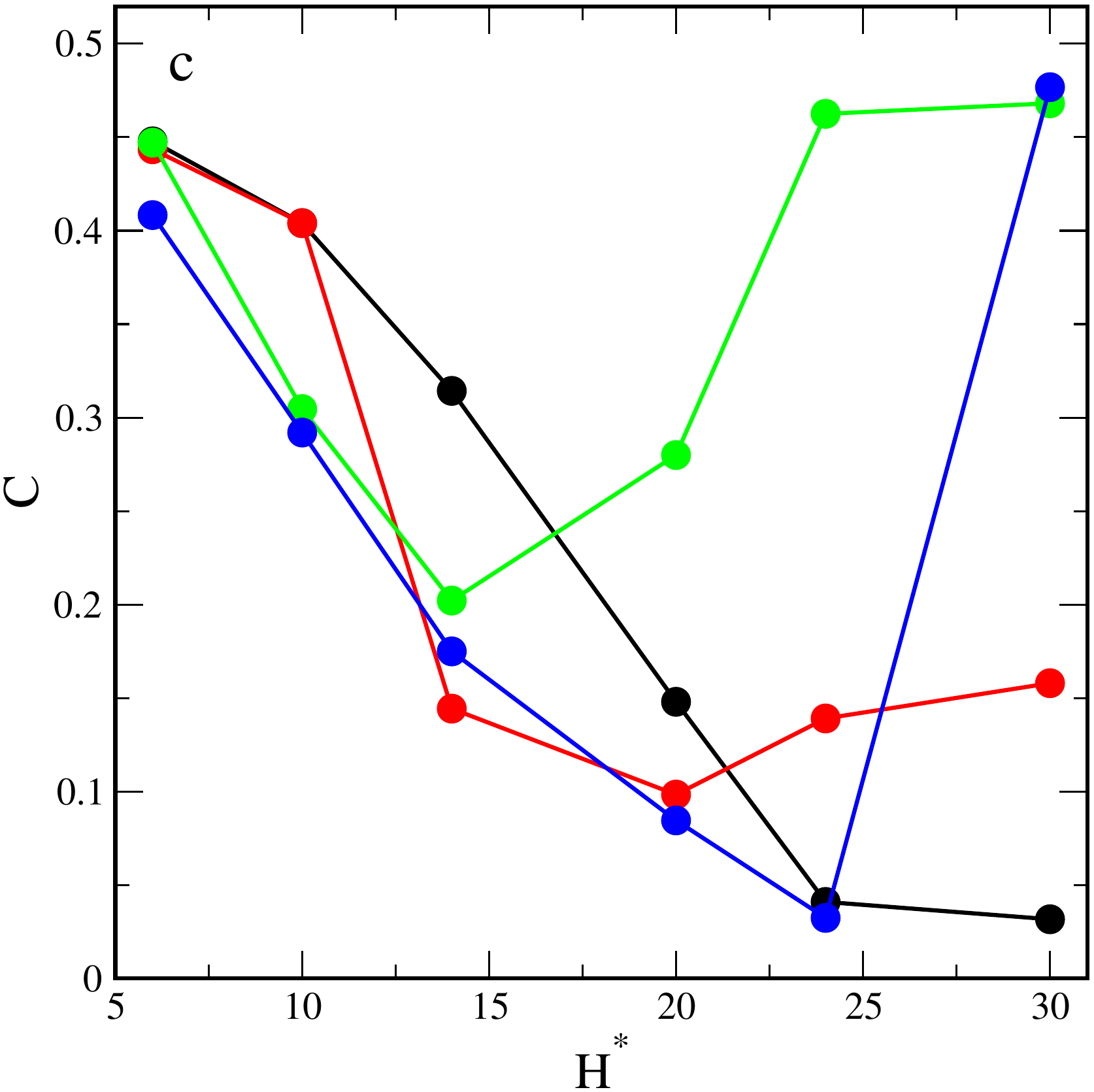} 
\caption{(Colour online) The relative shape asnisotropy (a), the prolateness (b) and the acillindricity (c) polotted as functions of the wall separation for repulsive walls (black lines) and attractive walls with different energy parameters $\varepsilon^*_s$: 1 (red lines), 3 (green lines), and 6 (blue lines).}
\label{fig5}
\end{figure}

The dependencies of shape parameters on the  width of the slit are presented in figure~\ref{fig5}. We begin with the analysis of the results obtained for inert walls. In this case, the relative shape anisotropy gradually decreases from 0.23 to a value close to zero, corresponding to spherical objects (part a). However, the prolateness increased from $-0.22$ to 0 (part b). This reflects the transformation from oblate structures under the strong confinement to spherical structures in wide pores. The acylindricity changes similarly to the $\kappa$. We see that the $C$ decreases from 0.45 to 0.03 (part c). Thus, there are no typically cylindrical structures. 

For attractive walls, as the pore width increases, $\kappa$ decreases to a minimum and increases again. If $\varepsilon_s=1$ or $\varepsilon_s=6$, the minima are at $H^*=20$, while for $\varepsilon_s=3$ it is at $H^*=14$. Then, an increase in the relative shape anisotropy is slight for the ``weak'' walls (M-structures) and rapid for the ``strong'' walls (M1-structures). The prolateness monotonously increases for the weakly attractive walls. However, for the strongly attractive surfaces, the $S$ initially increases to a maximum and rapidly decreases for wide slits. In the widest pores, flat symmetrical structures are formed at one of the walls. The acylindricity, however, has minimal values for the same $H^*$ as the relative shape anisotropy. It is interesting that for $H^*=20, 24$ and $\varepsilon_s=6$, the acylindricity is close to zero ($C=0.03$). Indeed, in this case, the spool with a long tube is found (see figure~\ref{fig3}d).

We see that the shape parameters quite well reflect the found particle conformations.

\subsection{The diagrams of a particle conformations}

\begin{figure}[!t] 
\centering
\includegraphics[width=9.0cm]{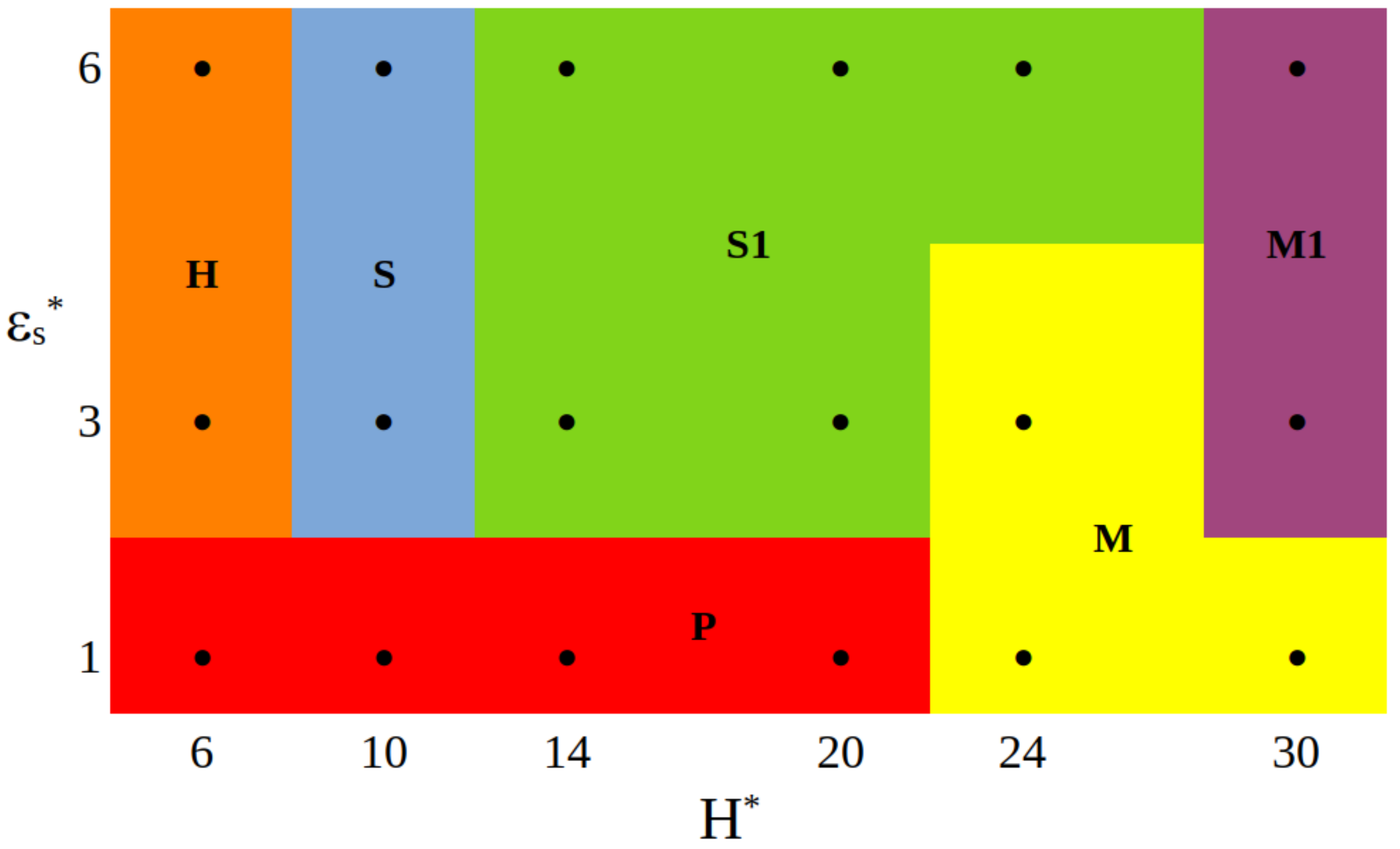}\\ 
\caption{(Colour online) The sketch diagram of the particle structures in the coordinates $\varepsilon_s^*$ vs $H^*$ for a hairy particle confined in a slit. Black circles correspond to simulation points. The boundaries between structures are arbitrarily chosen.}
\label{fig6}
\end{figure}

In figure~\ref{fig6}, we show what kind of structures are formed for different combinations of the strength of attractive segment-wall interactions and the wall separation. In the case of weakly attractive surfaces, the hairy particle has a pillar-like shape (P). However, for $H^*_s \geqslant 24$, the particle adsorbs on one surface and its canopy forms a mound (M). For ``stronger'' walls, however, the structure changes with an increase in the wall separation. In the case of narrow pores ($H^*_s \leq 24$), we see the following sequence of transitions: H $\to$ S $\to$ S1. For wider pores with $\varepsilon^*_s=3$, there is a gradual change of the structure: S1 $\to$ M $\to$ M1, while for $\varepsilon^*_s=6$, the asymmetrical spool transfers directly into the flattened mound M1.

\begin{figure} [!t]
\centering
\includegraphics[width=6.0cm]{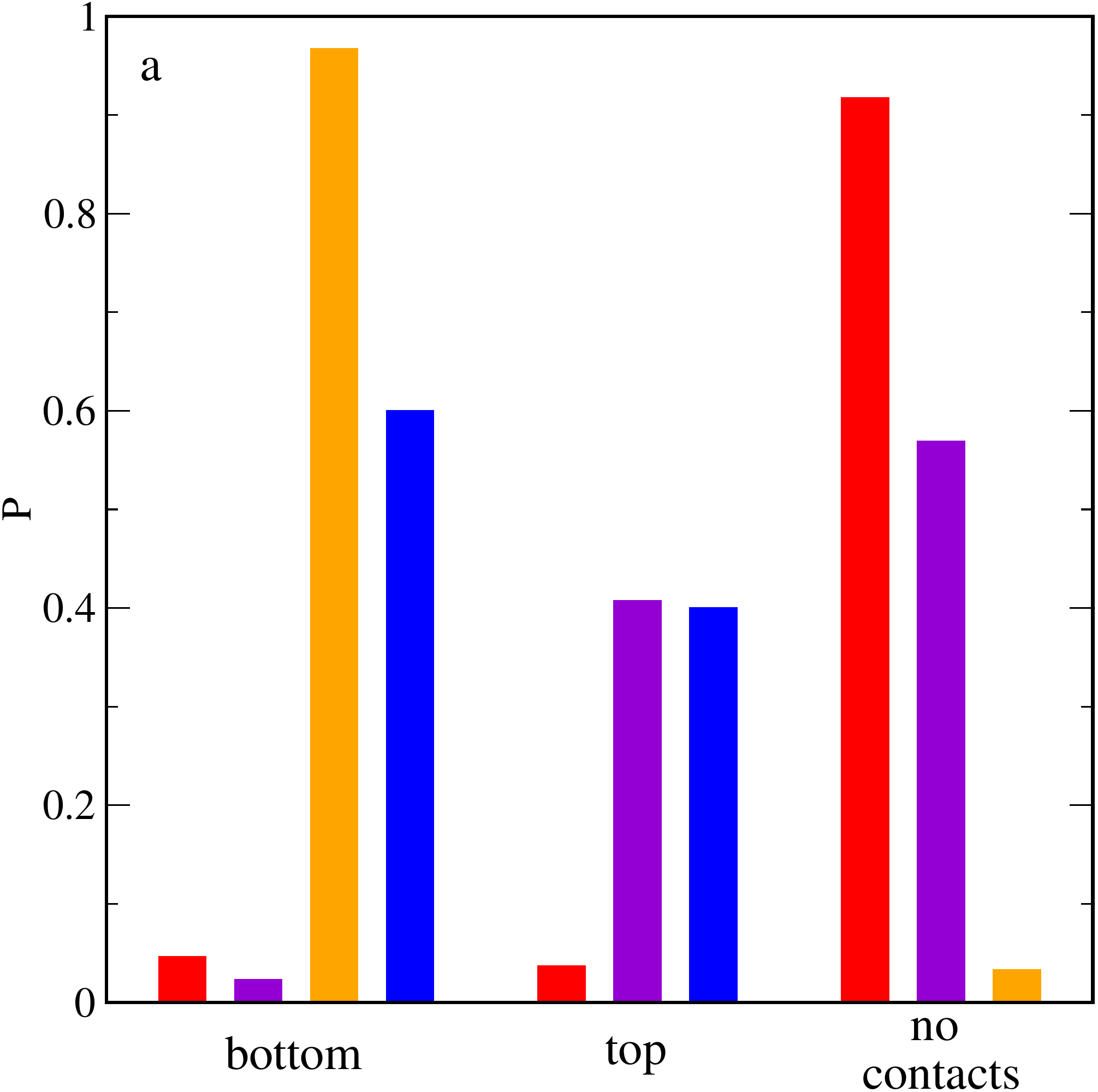} \includegraphics[width=6.0cm]{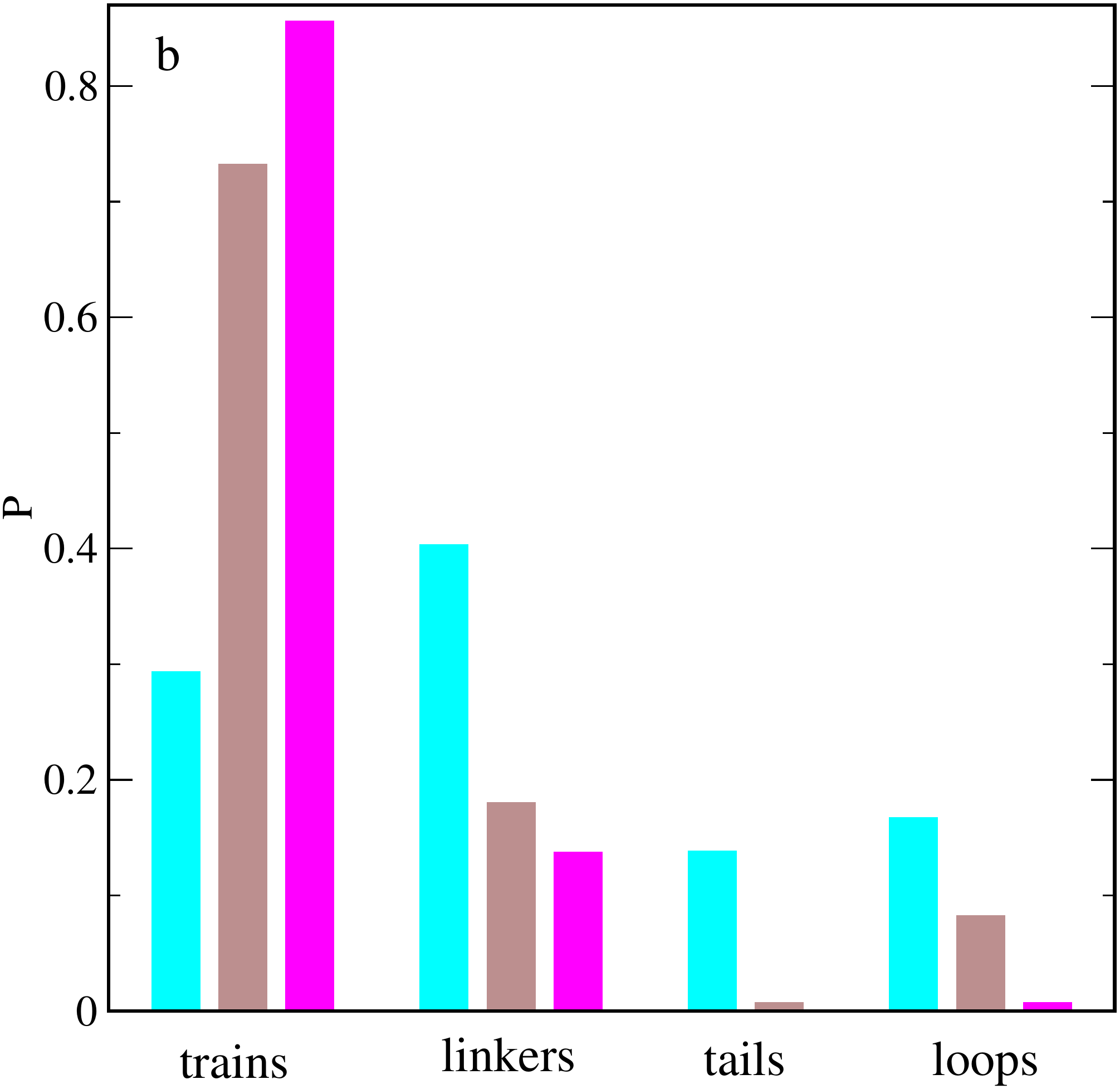} 
\caption{(Colour online) (a) Examples of distributions of chains in contact with the bottom wall, chains touching the top one, chains in contact with both walls and those without any contacts presented $H^*=24$ with repulsive walls (red bars) and attractive walls with $\varepsilon^*_s=1$ (violet bars), $\varepsilon^*_s=3$ (orange bars), and $\varepsilon^*_s$=6 (blue bars);
 (b) Examples of distributions of trains, loops, linkers and tails of chains adsorbed on the walls for $H^*=30$ and $\varepsilon^*_s=1$ (cyan bars), $\varepsilon^*_s=3$ (brown bars), and $\varepsilon^*_s$=6 (magenta bars).}
\label{fig7}
\end{figure}

To deeper understand the behavior of hairy particles between two walls, we compute the
fractions of chains being in contact with the bottom wall, touching the top wall, touching
both surfaces, and those without any contact with the walls. We assume that a given chain
is in contact with the wall if the distance from this wall, at least one segment of it, is less than
$1.5 \sigma$. The representative examples of these distributions obtained for different $\varepsilon^*_s$ are shown in figure~\ref{fig7}a. We see here that these distributions depend significantly on the strength of segment-wall interactions. For repulsive surfaces, almost all polymers do not touch any wall. In the case of weakly attractive walls, a mound-like structure is formed. Indeed, a lot of polymers  touch the bottom wall. However, the chains do not even come into contact with any wall. For $\varepsilon^*_s=3$, there are chains in contact with the bottom wall and those without any contacts (a lower mound). In the case of $\varepsilon^*_s=6$, a spool-like structure is observed. Therefore, the polymers only touch one wall or the other.

We have also monitored the configurations of chains touching a given surface. We distinguish
four types of chain fragments, the standard structures, namely trains, loops, tails,
and a new structure, linkers. The linker is a tail attached to the core. Figure~\ref{fig7}b shows these
distributions obtained for $H^*=30$ and different values of $\varepsilon^*_s$. As the strength of segment-wall interaction increases, the fraction of trains also increases but the opposite effect is visible for linkers, tails, and loops. In the case of $\varepsilon^*_s=6$, there are no tails at all (starfish-like structure \cite{28}). 

\section{Conclusions}

We performed simulations for the model ligand-tethered particles confined between two inert or attractive walls. We assumed that all interactions but the segment-wall ones were softly repulsive, and discussed the influence of segment-wall interactions and the wall separation on the equilibrium configurations of hairy particles. 
 
In the case of inert walls, the hairy particles are spheroids, which become more and more flattened as the pore width increases. The most interesting findings are those obtained for attractive walls. We found here two basic conformations of hairy particles: bridges and mounds. The particle canopy can form a bridge between the walls. So far such bridges were not studied. 

The bridges are similar to flanged spools. We observed four spool-like conformations: symmetrical spools (S), asymmetrical spools (S1), hourglass (H), and pillars (P). For the symmetric spool, the core is located at the center of the slit. Moreover, the ``flanges'' are almost the same, However, for asymmetrical spools, the core lies closer to one of the walls and the ``collars'' are different.  In the case of the hourglasses, the density of segments is very low in the central part of the pore and gradually increases reaching the maxima at walls. For pillars, the segment density is only slightly higher near the walls than in the center.  

Under certain conditions, the particles fall on one of the walls and resemble mounds. For weakly attractive walls, relatively high mounds are formed (M). However, for strongly attractive surfaces, we found flattened mounds (M1). The latter structures are quite similar to the starfish configuration of adsorbed star polymers, where all chains collapsed onto the surface~\cite{28}. The structures formed on a single solid surface were observed experimentally \cite{23} and were predicted by computer simulations~\cite{25, 27}. 

We present an overview of the observed structures in the schematic diagram in the coordinates~$H^* ~-~ \varepsilon^*$. 
 
We also show how the wall separation and the strength of segment-wall affect the parameter describing the shapes of segment clouds, the radius of gyration, its components in the Cartesian coordinates, the prolateness, the relative shape anisotropy, and acylindricity. 

We discuss here the behavior of an isolated hairy particle confined between attractive walls. However, this work is a starting point for the study of the self-assembly of hairy particles between two planes and the adsorption of hairy particles in porous materials.

In summary, we have proved that confinement in attractive slits can be a method for the control of the shape of hairy particles.



\ukrainianpart

\title[Полімерна частинка у щілині]%
{Полімерна частинка у щілині}%

\author[T. Сташевський, M. Борувко]{T. Сташевський, M. Борувко}

\address{Кафедра теоретичної хімії, Інститут хімічних наук, хімічний факультет, університет ім. Марії Склодовської-Кюрі, Люблін, Польща}

\makeukrtitle

\begin{abstract}

Методом молекулярної динаміки досліджуються зміни форми щіткоподібних наночастинок
у обмеженому просторі.
Обговорюється поведінка таких частинок в щілинах з інертними
або притягувальними стінками. Припускається, що лише взаємодії між 
ланцюжками та стінками є притягувальними. Досліджено залежність конфігурації 
частинок від сили взаємодії зі стінками та ширини щілини. Для випадку притягувальних 
поверхонь знайдені нові структури, в яких ланцюжки пов'язані з обидвома стінками та 
утворюють містки між ними: колони, симетричні й асиметричні каркаси та ``пісочні 
годинники''. У широких порах з сильно притягувальними стінками щіткоподібні 
частинки адсорбуються на одній з поверхонь та утворюють ``горбочки'' або ж структури 
типу морської зірки.

	\keywords щіткоподібні частинки, щілиноподібні пори, молекулярна динаміка
\end{abstract}

 \end{document}